\documentclass[aps,prstab,preprint,showpacs,groupedaddress]{revtex4}
\usepackage{graphics,epsfig}
\begin{document}

\draft
\title{Effects of Contrarians in the Minority Game}

\author{Li-Xin Zhong$^{1}$, Da-Fang Zheng$^{1,}$\footnote{Corresponding author.\\ E-mail address:
dfzheng@zjuem.zju.edu.cn.}, Bo Zheng$^{1}$, and P.M. Hui$^{2}$}

\affiliation{$^{1}$ Zhejiang Institute of Modern Physics, Zhejiang
University, Hangzhou 310027, People's Republic of China \\
$^{2}$ Department of Physics, The Chinese University of Hong Kong,
Shatin, Hong Kong, China}

\date{\today}

\begin{abstract}
We study the effects of the presence of contrarians in an
agent-based model of competing populations.  Contrarians are
common in societies.  These contrarians are agents who
deliberately prefer to hold an opinion that is contrary to the
prevailing idea of the commons or normal agents.  Contrarians are
introduced within the context of the Minority Game (MG), which is
a binary model for an evolving and adaptive population of agents
competing for a limited resource.  Results of numerical
simulations reveal that the average success rate among the agents
depends non-monotonically on the fraction $a_{c}$ of contrarians.
For small $a_{c}$, the contrarians systematically outperform the
normal agents by avoiding the crowd effect and enhance the overall
success rate. For high $a_{c}$, the anti-persistent nature of the
MG is disturbed and the few normal agents outperform the
contrarians. Qualitative discussion and analytic results for the
small $a_{c}$ and high $a_{c}$ regimes are also presented, and the
crossover behavior between the two regimes is discussed.

\end{abstract}


\maketitle

\section{Introduction}
\label{sec:introduction}

In real-life societies, there always exist some people, referred
to as ``contrarians", who deliberately prefer to take on an
opinion contradictory to the prevailing thoughts of others.
Contrarian investment strategies, for example, have been an active
and important subject of studies in finances \cite{dreman,corcos}.
Besides financial markets, effects of the existence of contrarians
have also been studied recently within the context of the dynamics
of opinion formation in social systems \cite{galam, mobilia}.  In
the present work, we explore how contrarians may affect the global
features in one of the most popular agent-based models in recent
year, namely the Minority Game (MG) \cite{challet}. The MG is a
binary version of the type of problems related to the
bar-attendance problem proposed by Arthur \cite{arthur}. It has
become the basic model of competing populations with built-in
adaptive behavior \cite{web}. On the application side, it has been
shown that the MG can be suitably generalized to model financial
markets and reproduce the so-called stylized facts observed in
real markets \cite{challet1,johnson}.

In the basic MG, an odd number $N$ of agents decide between two
possible choices, say, 0 or 1, at each timestep. The winners are
those belonging to the minority group. The winning outcome can,
therefore, be represented by a single digit: 0 or 1, according to
the winning option.  The most recent $m$ winning outcomes
constitute the only information that is made available to all
agents.  The agents decide based on this global information. For
given $m$, there are $2^{m}$ possible $m$-bit history bit-strings,
leading to a strategy space consisting of a total of $2^{2^{m}}$
possible strategies.  Each strategy gives a prediction of either 1
or 0 for each of the $2^{m}$ histories. Initially, each agent
picks $s$ strategies at random, with repetitions allowed. The
performance of the strategies is recorded by assigning (deducting)
one (virtual) point to the strategies which would have predicted
the correct (incorrect) outcome, after the outcome is known in a
timestep.  At each timestep, each agent follows the prediction of
the momentarily best-performing strategy, i.e., the one with the
highest virtual points among her $s$ strategies.  Therefore, a
feedback mechanism is built in by allowing the agents to adapt to
past performance, which in turn is related to the actions of the
agents themselves, by shifting from one strategy to another.

Despite the simplicity of the model, the MG shows very rich and
non-trivial properties \cite{savit,challet2,zheng}. A quantity of
interest, for example, is the standard deviation $\sigma$ in the
number of agents making a particular decision, averaged over
different runs. This quantity characterizes the collective
efficiency of the system in that a small $\sigma$ implies a higher
success rate or winning probability per agent and hence more
winners per turn. Most noticeably, $\sigma$ exhibits a
non-monotonic behavior on $m$ \cite{savit,challet3,manuca},
showing a minimum at which the performance of the system is better
than that of a system in which the agents decide randomly.  This
feature can readily be explained in terms of the crowd-anticrowd
theory of Johnson and coworkers \cite{johnson1,hart}. If the
number of strategies in the whole strategy pool is smaller than
the total number of strategies in play, many agents will hold
identical strategies. With decisions based on the best-performing
strategies, many agents will then make identical decisions and
form a ``crowd".  For small $m$, $\sigma$ is large due to the lack
of a cancellation effect from a corresponding ``anti-crowd" of
agents using the opposite or anti-correlated strategy.  This small
$m$ regime is referred to as the informationally efficient phase,
as there is no information in the history bit-strings that the
agents can exploit \cite{savit}. In the large $m$ limit, however,
the strategy pool is much larger than the number of strategies in
play. Thus it is unlikely that a strategy is being used by more
than one agent, and the best performing strategies are those not
in play among the agents. In this case, the agents behave as if
they are deciding independently and randomly, leading to
$\sigma\sim\sqrt{N}/2$.  The minimum value of $\sigma$ occurs at
around $2 \cdot 2^{m} \sim Ns$, where the size of the crowd and
anti-crowd become comparable \cite{johnson1,hart}.

An important and interesting question is whether the collective
efficiency in MG can be optimized and how, especially in the
efficient regime of the system.  A few modified versions of the MG
have been proposed and studied, with an enhanced performance to
different extent \cite{cavagna,johnson2,cara,li}. In the present
work, we show that a population consisting of a small fraction of
agents with contrarian character, i.e., agents who act opposite to
the prediction of their own best-performing strategies, will have
the standard deviation highly suppressed and hence the performance
of the population greatly improved.  The plan of the paper is as
follows. In Sec. \ref{sec:model}, we present our model and define
the action of the contrarians. In Sec. \ref{sec:results}, we
present results of extensive numerical simulations, together with
qualitative discussion and analytic results on the behavior in the
limits of small and large fractions of contrarians.  The crossover
behavior between the two limits is also discussed. Section
\ref{sec:summary} summarizes the present work.

\section{Minority game with contrarians}
\label{sec:model}

We consider a system of an odd number $N$ of agents, including
$N_{n}$ {\it normal} agents and $N_{c}$ {\it contrarians},
competing for a limited resource. At each timestep, each agent
must choose one of two options, $0$ or $1$. The winners are those
belonging to the minority group and a winner is awarded one (real)
point for each winning action.  The only information available to
all agents is the history bit-strings recording the most recent
$m$ winning outcomes (i.e., the minority sides).  For a given
value of $m$, there are $2^{m}$ possible histories.  A strategy
gives a prediction for each of the $2^{m}$ histories, and
therefore the whole strategy pool has $2^{2^{m}}$ strategies in
total. Initially, each agent randomly picks $s$ strategies from
the strategy pool, with repetitions allowed. After each timestep,
each strategy is assessed for its performance by rewarding one
(virtual) point (VP) to the strategy if it has predicted the
winning option.  The VPs thus reflect the cumulative performance
of the strategies that an agent holds from the beginning of a run.
At each timestep, each agent makes use of the momentarily
best-performing strategy, i.e., the one with the highest VP, in
her procession for decision.  A random tie-breaking rule is used
in case of tied VPs.  However, normal agents and contrarians use
the best-performing strategies in {\em different} ways.  For a
normal agent, she {\em follows} the predictions of the
best-performing strategy. For a contrarian, however, she takes the
opposite (hence the name contrarian) action to the prediction of
her best-performing strategies, i.e., if the strategy with the
highest VP says 0, for example, a contrarian will choose option
$1$.  Note that the assignment of VPs to strategies does {\em not}
depend on the type of agents under consideration.  For a
contrarian, for example, if she loses in a timestep, her
best-performing strategy has actually predicted the correct
outcome and hence a VP will be rewarded.

\section{Results and Discussions}
\label{sec:results}

We have performed extensive numerical simulations to study the
effects of the presence of contrarians in MG.  Typically, we
consider systems of $N=101$ agents, with $s=2$ strategies per
agent.  Each run lasts for $10^{4}$ timesteps and each data point
represents an average over the results of $50$ independent runs of
different initial distributions of strategies and initial
histories in starting the runs.  Figure 1(a) shows the averaged
success rate $R$ over all the agents, which is the number of real
points per agent per turn, as a function of the parameter $m$, for
different fractions of contrarians $a_{c}=N_{c}/N$ in the system.
For $a_{c}=0$, our model reduces to the MG. The results show that,
for small $m$ corresponding to the efficient phase of the basic
MG, and for small $a_{c}$, $R$ increases as $a_{c}$ increases. For
about $a_{c}=0.2$-$0.3$, $R$ achieves a maximum value of about
0.485 at $m=4$. Note that by definition the highest value of $R$
is $(N-1)/2N$. The effects of contrarians are also reflected in
the averaged standard deviation $\sigma$ (see Figure 1(b)), which
drops sensitively with $a_{c}$ for small values of $a_{c}$ in the
efficient phase.  In the absence of contrarians, $\sigma$ shows a
minimum at about $m=6$ for the size of system considered
\cite{savit}.  In the presence of contrarians,  the minimum in
$\sigma$ now occurs at $m=4$, with a suppressed value of $\sigma$
when compared with the basic MG. The success rate and $\sigma$ are
related in that a smaller $\sigma$ implies more winners per turn
and hence a larger success rate \cite{johnson3}. The results in
Fig. 1, therefore, indicate that the presence of a small fraction
of contrarians enhances the collective performance of the system
in the efficient phase (small $m$) of the MG.

As the key feature of improved performance occurs at small values
of $m$, we will focus in this regime from now on. Figure 2
summarizes the dependence of the averaged success rate $R$ on the
fraction of contrarians $a_{c}$, for $m=2,3,4$.  For small
$a_{c}$, $R$ increases sensitively with $a_{c}$, and achieves a
maximum at some value of $a_{c}$. We have checked that this value
of $a_{c}$ tends to increase towards $1/2$ as the number of agents
$N$ in the system increases.  The maximum value of $R$ and the
fraction $a_{c}$ for achieving the maximal $R$ are both
$m$-dependent.  An overall maximal value of $R \approx 0.485$
occurs at $m=4$ and $a_{c} \approx 0.25$-$0.3$, for systems with
$N=101$. For large $a_c$, $R$ decreases as $a_{c}$ increases and
the corresponding values of $R$ become less sensitive to $m$. It
is also interesting to investigate the success rates averaged over
the normal agents and averaged over the contrarians separately, to
see how the averaged results in Fig.2 comes about. Figure 3 shows
the results for $m=2$, which are typical of the small $m$ cases
shown in Fig.2. The results indicate that a small fraction of
contrarians can {\em systematically} take advantage of the
background normal agents, as the contrarians have a success rate
that is significantly higher than $1/2$, while the normal agents
basically take on a constant success rate corresponding to that of
the basic MG with the same value of $m$.  However, as the fraction
of contrarians becomes large, it is the remaining few normal
agents who take advantage of the contrarians and attain a success
rate of about $0.7$, while the contrarians only have a success
rate of about $0.25$.  A crossover between these two regimes
occurs at an intermediate fraction of $a_{c} \approx 0.36$ for
$m=2$ and $N=101$ where the success rates of the two types of
agents are comparable.

The behavior for a small fraction of contrarians can be readily
understood within the physically transparent crowd-anticrowd
picture of the MG \cite{johnson1,hart}. For a small fraction of
contrarians, the winning outcomes and hence the strategies's VPs
are still dominated by the behavior of the normal agents.
Therefore, the behavior of the system basically follows that of
the basic MG.  For the present model, the most important point to
realize is the anti-persistent nature of the system, i.e., there
is no runaway VPs for the strategies \cite{zheng}.  In other
words, strategies that have predicted the correct (incorrect)
outcomes in recent turns are bound to predict incorrectly
(correctly) in future turns.  This leads to the so-called doubly
periodic behavior in the outcomes \cite{zheng} as it takes the
system about $2\cdot 2^{m}$ timesteps to pass through an Eulerean
trail in the history space formed by all the $2^{m}$ possible
histories \cite{lo,sonic}. For small $m$, the number of strategies
in play is larger than the total number of strategies, implying an
appreciate overlap of strategies among the agents. The lower
success rate of the normal agents (see Fig.3) comes about from the
crowd effect, i.e., a group of agents using the same or similar
better-performing strategies for decision at a timestep. For small
$m$, this crowd is too big to win.  A low success rate (relative
to $(N-1)/2N$) or a large $\sigma$ implies that there is a room
for more winners per turn. A contrarian, by taking the opposite
action of the prediction of the best-performing strategy that she
holds, is given the ability to avoid herself from joining the
crowd and win more frequently than the minority rule allows. This
breaking away from the crowd has the effect of allowing more
winners per turn and hence suppressing the standard deviation (see
Fig.1(b)). It is worth noting that several variations of the MG
also give an enhanced success rate under some condition. For
example, the presence of a fraction of agents who decide based on
a larger value of $m$ in a background of agents using a smaller
value of $m$ also gives rise to an enhanced overall success rate
\cite{johnson3}. Another way of breaking away from the crowd is to
allow some agents to opt out of a MG at random timesteps
\cite{larry}.  Quantitatively, one expects that in the large $N$
limit, since the $N(1-a_{c})$ normal agents have a success rate $R
= R(a_{c}=0) \equiv R(0)$ and the outcomes are dominated by the
normal agents for small $a_{c}$, the $Na_{c}$ contrarians have a
success rate of $1-R(0)$. Averaging over the normal and contrarian
agents gives a success rate $R(a_{c}) = R(0) + a_{c} (1 - 2R(0))$,
which is a good approximation of the numerical results in Fig.2
for small $a_{c}$.  Note that $R(0)$ is $m$-dependent, due to the
better crowd-anticrowd cancellation effect as $m$ increases.  The
value of $R(0)$ can be obtained by invoking the analytic
expressions within the crowd-anticrowd theory
\cite{johnson1,hart}.  For our purpose, it is sufficient to take
the value of $R(0)$ from the numerical results at $a_{c}=0$.

As the fraction of contrarians increases, the features in the
winning outcome series start to deviate from that of the basic MG
(i.e., without the contrarians).  In particular, the
anti-persistence of the strategies' VPs may be destroyed. This
will lead to some strategies with runaway VPs, i.e., VPs that keep
on increasing or decreasing in a run.  Accompanying this effect is
the emergence of biased conditional probabilities in the winning
outcome series. In the basic MG, the probability of having a
winning outcome of 1 following a given history of $k$ bits ($k
\leq m$) is equal to that of having a winning outcome of 0
following the same $k$-bit history \cite{savit}.  To illustrate
the change in this basic feature of the MG as the fraction of
contrarians increases, we study the quantity \cite{challet1}
\begin{equation}
H = \frac{1}{Q} \sum_{\mu} [  P(0|\mu) - P(1|\mu)]^{2},
\end{equation}
where the sum is over all the $Q$ possible histories $\mu$ of a
certain bit-length and $P(i|\mu)$ is the conditional probability
of having a winning outcome of $i$ ($i$=0 or 1) given the history
bit-string is $\mu$.  For the basic MG, $H=0$ as the two
conditional probabilities cancel.  We, therefore, expect that as
$a_{c}$ increases, $H=0$ in a range of $a_{c}$ for which the
contrarians are too few to affect the outcomes but can efficiently
avoid the crowd effect. For large $a_{c}$, $H>0$ as the system
becomes increasingly deviated from the anti-persistent behavior.
Figure 4 shows the dependence of $H$ for histories of $m$-bits on
the fraction of contrarians, for systems with $m=2,3,4$.
Interestingly, the range of $a_{c}$ with $H=0$ corresponds to the
same range that the success rate of the normal agents is flat (see
Fig.~3). The results indicate that the winning outcomes series has
similar features as in the basic MG in this range of small
$a_{c}$, hence justifying our previous discussion on the small
$a_{c}$ behavior. As $a_{c}$ further increases, $H$ starts to
deviate from zero at a $m$-dependent value of $a_{c}$ for fixed
$N$. This value of $a_{c}$ increases towards $1/2$ for larger $N$.
The result indicates that the contrarians are not only simply
adapting to the actions of the normal agents, but also affecting
the outcomes themselves.

For sufficiently high $a_{c}$, $H=1$, indicating a highly biased
conditional probability. Numerically we have checked that in many
runs at high $a_{c}$, the system shows a persistent outcome (of
1's or 0's).  For these runs, it is expected $(1 - 1/2^{s})N =
3N/4$ agents {\em hold} a strategy that predicts the persistent
winning outcome regardless of the value of $m$ in the large $N$
limit.  It is because the system is now restricted to a tiny
portion of the history space and it is the prediction based only
on {\em one} particular history bit-string (out of $2^{m}$) in a
strategy that really matters \cite{sonic}. Due to the minority
rule, an outcome series of persistent winning option is not
allowed in the absence of contrarians.  In the presence of a large
fraction of contrarians in the $3N/4$ agents holding strategies
with runaway VPs, however, persistent winning outcomes are
allowed.  It is because the contrarians take the opposite action
to the prediction of the strategies and leave the room for the
normal agents to become winners. Among the few $N(1-a_{c})$ normal
agents, $3/4$ of them win persistently and $1/4$ of them lose
persistently. This gives rise to the high averaged success rate
(nearly $0.7$) for the normal agents at high $a_{c}$ (see Fig.3).
For the $Na_{c}$ contrarians, $1/4$ of them persistently take the
winning action as they hold a strategy that persistently predicts
a wrong outcome. This gives rise to the averaged success rate of
about $0.25$ for the contrarians at high $a_{c}$, as shown in
Fig.3.  If we consider the number of winners collectively, there
are $\frac{3N}{4} (1 - a_{c})$ winners from the normal agents and
$\frac{N}{4} a_{c}$ winners from the contrarians per turn.  This
leads to an overall success rate of $R(a_{c}) = (\frac{3}{4} -
\frac{a_{c}}{2})$ for sufficiently high $a_{c}$. From Figs.3 and
4, we notice that $a_{c} > 0.6$ corresponds to the high $a_{c}$
regime.  The result gives a continuous drop of $R$ as $a_{c}$
increases towards unity, as observed in Fig.2. For $a_{c}=1$, $R =
1/4$ for runs with persistent winning outcomes. Note that our
argument does not depend on the value of $m$, as long as the
number of agents is sufficiently large.  In Fig.2, we observed
that the results for $m=2,3,4$ become less sensitive to $m$ in the
high $a_{c}$ regime, as predicted. The $m=2$ results follow our
prediction reasonably well. The discrepancies from the prediction
in the $m=3$ and $4$ results come from the small size of system
($N=101$) that we used in the numerical simulations, and the fact
that there are runs for which the outcome series is different from
a persistent winning option. For example, it is possible to have a
series consisting of alternating winning options in the high
$a_{c}$ regime. However, the discussion based on outcomes with
persistent winning option does capture the essential underlying
physics embedded in the numerical results.

In Fig.2, $R(a_{c})$ shows a peak at a $m$-dependent crossover
value $\overline{a_{c}}$.  This value is also found to increase
towards $1/2$ as $N$ increases.  Following our discussions on the
small $a_{c}$ and high $a_{c}$ regimes, we may estimate
$\overline{a_{c}}$ by approximating it to be the value of $a_{c}$
at which the small $a_{c}$ behavior crosses over to the high
$a_{c}$ behavior, without considering the details of the
intermediate regime.  Thus, $\overline{a_{c}}$ can be determined
by $R(0) + \overline{a_{c}} (1 - 2R(0)) = 3/4 -
\overline{a_{c}}/2$, giving $\overline{a_{c}} = 1/2$ in the limit
of large $N$.  The $m$-dependence on $\overline{a_{c}}$ comes from
the finite size of $N=101$ agents in the systems that we used in
numerical simulations.  For finite $N$, the rapid increase in the
strategy pool size leads to a rapid drop in overlap of strategies
among the agents as $m$ increases.  This has the effects of
suppressing the range of $a_{c}$ in which the behavior in the
small $a_{c}$ regime is valid and enlarging the range of
intermediate $a_{c}$ where the system takes on complicated paths
in the history space other than the Eulerean trail in small
$a_{c}$ and the highly restricted path to a small portion of the
history space at high $a_{c}$.

\section{Summary}
\label{sec:summary}

We proposed and studied a generalized minority game consisting of
a fraction of contrarians.  These contrarians prefer to hold an
opposite opinion to the commons or the normal agents. Within the
context of MG, the contrarians are assumed to always take the
opposite action as predicted by their momentarily best-performing
strategy.  For small fraction of contrarians, the winning outcomes
are dominated by the normal agents and the contrarians can
systematically outperform the normal agents by avoiding the crowd
effect and hence the losing turns of the normal agents.  This
leads to an enhanced overall success rate of the system at small
$a_{c}$.  However, a larger fraction of contrarians will alter the
features in the outcome winning series, as compared to the basic
MG.  The results indicate that at high fraction of contrarians,
the few normal agents have a substantively higher success rate
than the contrarians. This is related to the change from
anti-persistent to runaway behavior in strategy performance, as
$a_{c}$ increases.  This change in character leads to a
non-monotonic dependence of the average success rate among all
agents as a function of $a_{c}$.  The small $a_{c}$ behavior can
be understood within the crowd effect in MG and the high $a_{c}$
behavior is dominated by the runs with persistent winning
outcomes. Analytic expressions were given for both the small
$a_{c}$ and high $a_{c}$ regimes.  In the limit of large $N$, it
is expected that the crossover between the two regimes occurs at
$\overline{a_{c}} = 1/2$.

\section*{ACKOWLEDGMENTS}
This work was supported by the National Natural Science Foundation
of China under Grant Nos. 70471081, 70371069, and 10325520. One of
us (P.M.H.) acknowledges the support from the Research Grants
Council of the Hong Kong SAR Government under Grant No.
CUHK4241/01P.


\newpage \centerline{\bf FIGURE CAPTIONS}

\bigskip
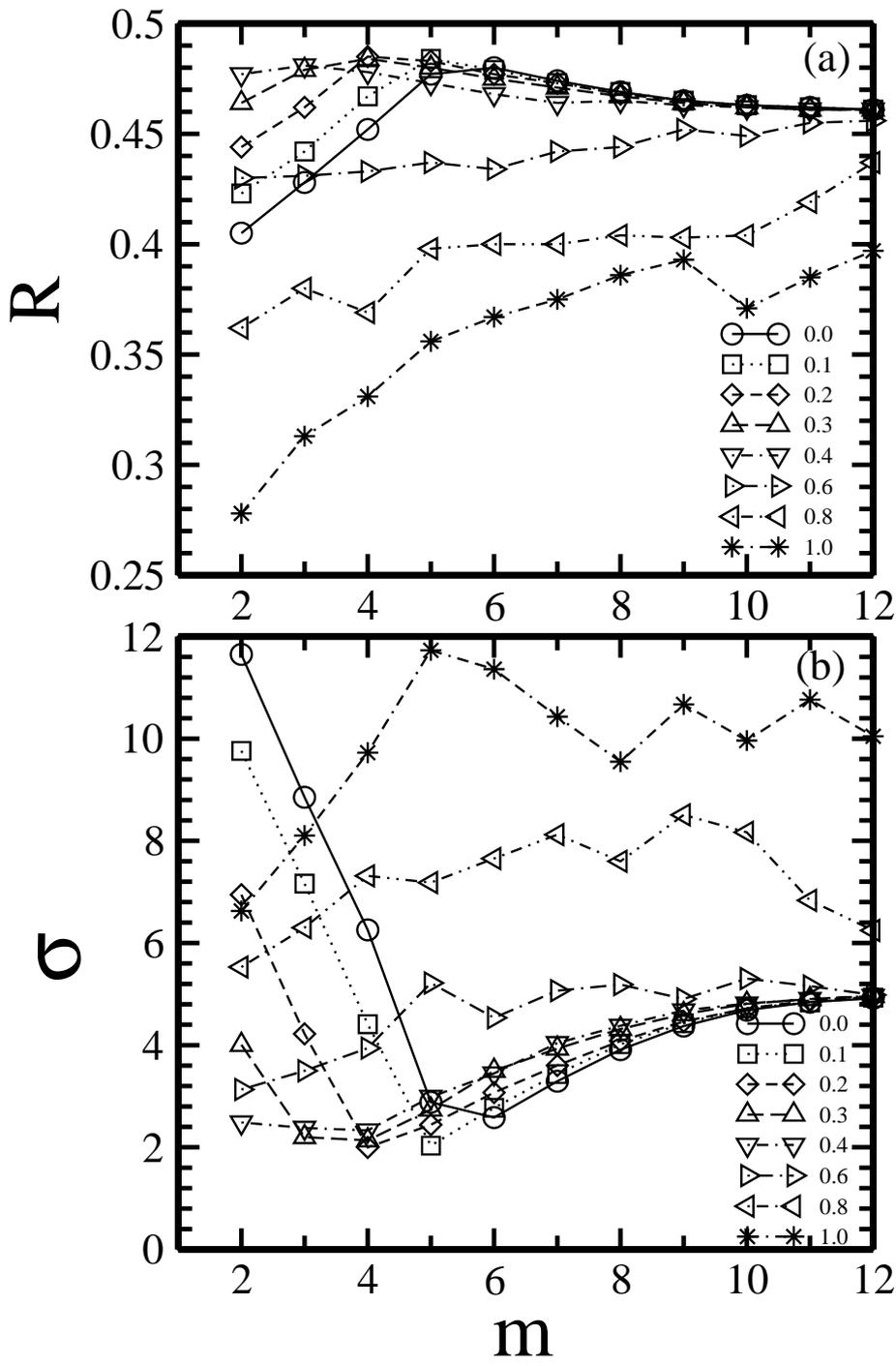
\noindent Figure 1: (a) The averaged success rate $R$ and (b) the
averaged standard deviation $\sigma$, as a function of the history
bit-length $m$ based on which agents decide.  Numerical results
for different fractions of contrarians $a_{c}=0.0, 0.1, 0.2, 0.3,
0.4,0.6, 0.8, 1.0$ are shown in different symbols.

\bigskip
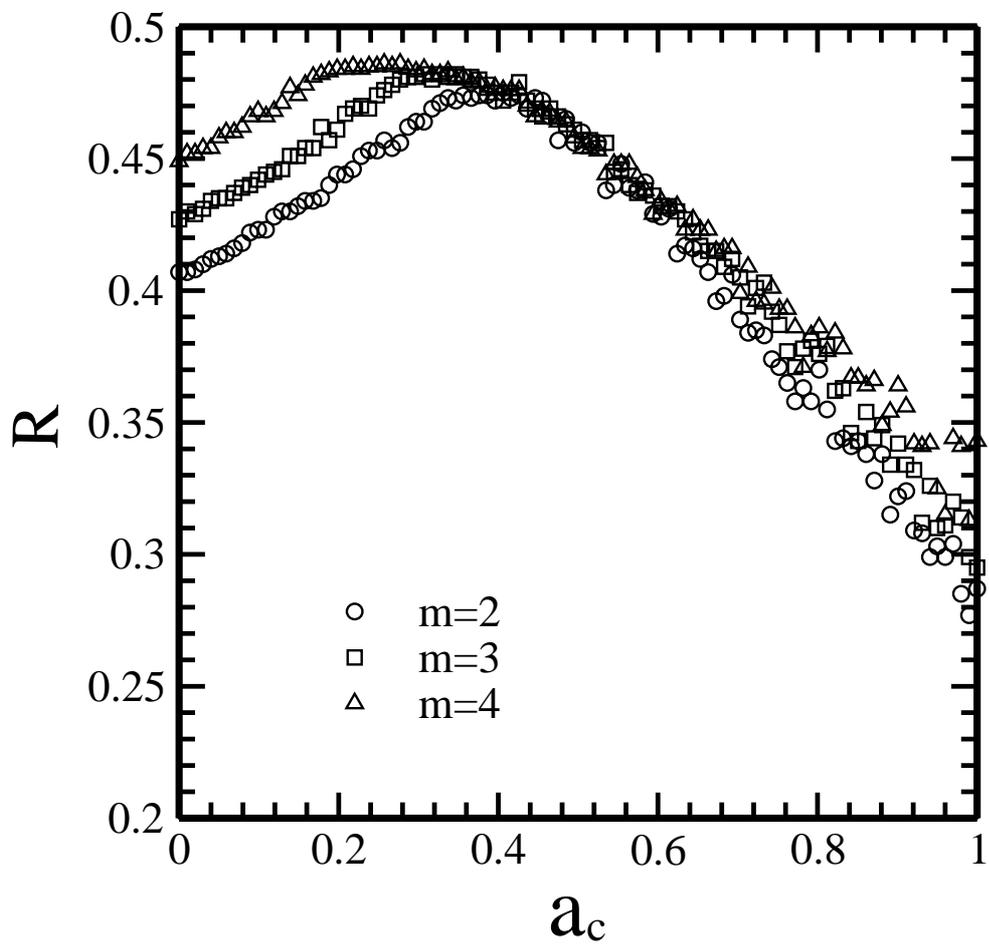
\noindent Figure 2: The average success rate $R$ as a function of
the fraction of contrarians $a_{c}$, for different values of
$m=2$, $3$, and $4$.

\bigskip
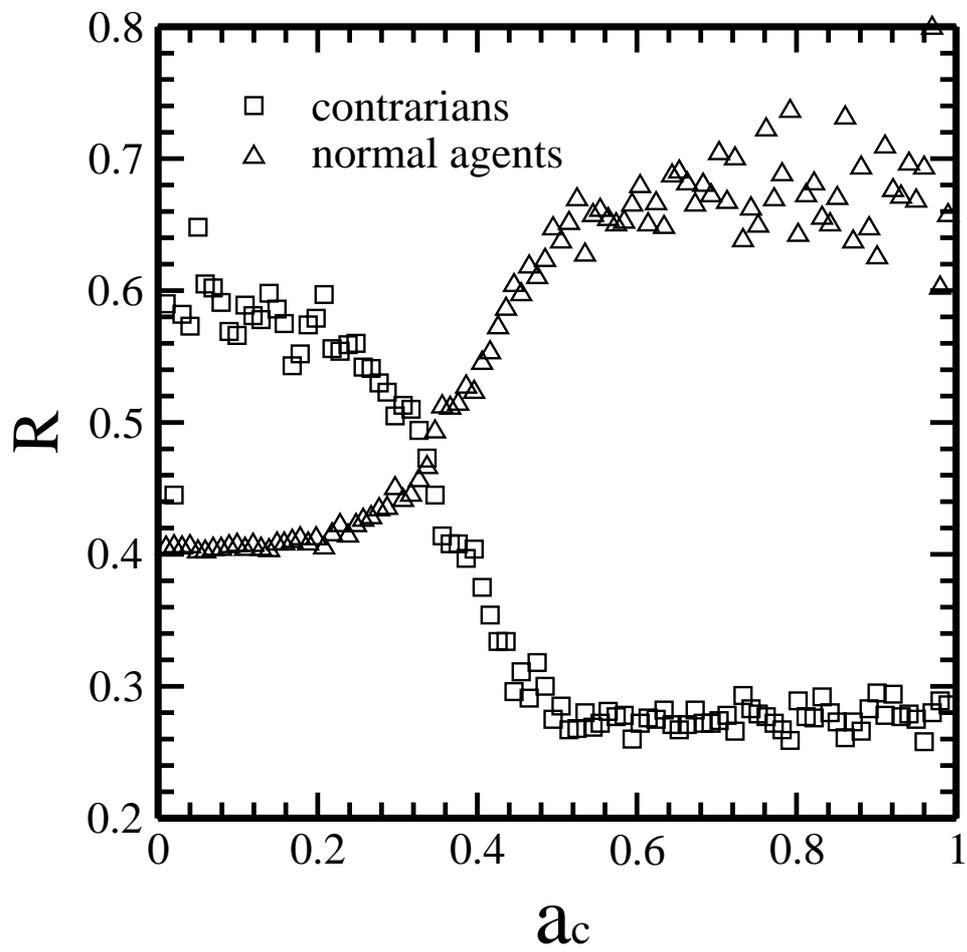
\noindent Figure 3: The averaged success rates $R$ of the normal
agents and of the contrarians as a function of the fraction of
contrarians $a_{c}$.  The results are for the case of $m=2$.

\bigskip
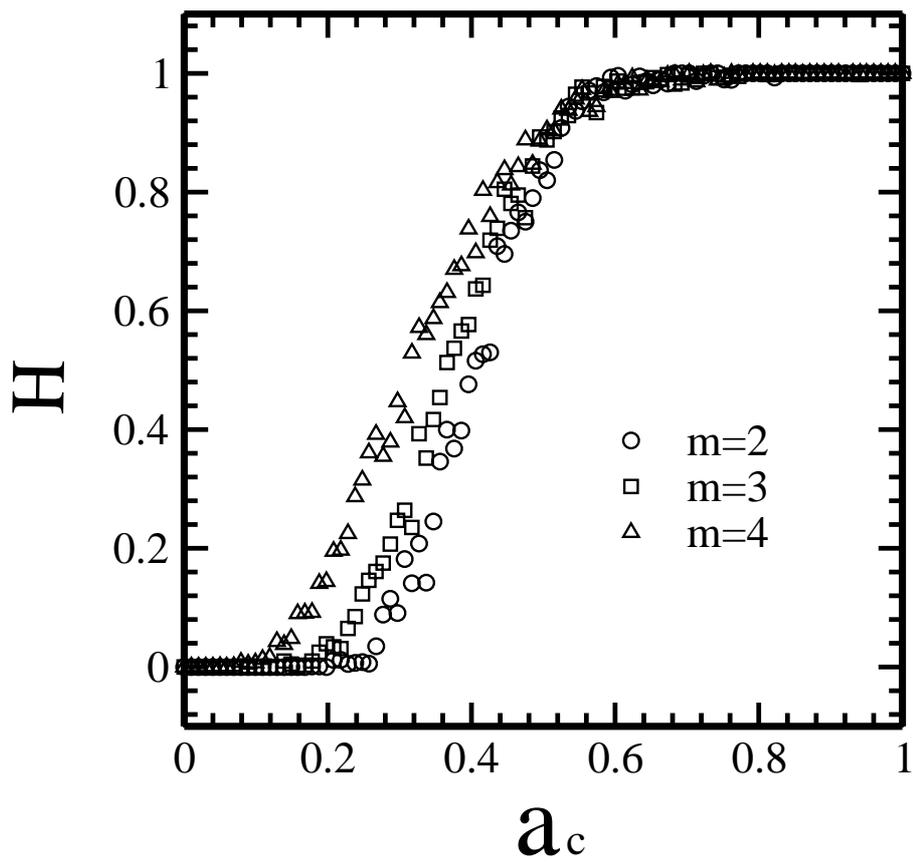
\noindent Figure 4: The quantity $H$ as a function of the fraction
of contrarians $a_{c}$ for $m=2$, $3$, and $4$.  For a range of
small $a_{c}$, $H=0$ showing that the outcome winning series is
dominated by the normal agents.  For high $a_{c}$, $H=1$
indicating that the anti-persistent nature of the system has been
disturbed by the presence of contrarians.


\begin{thebibliography}{99}

\bibitem{dreman} D. Dreman,
            {\it Contrarian Investment Strategies} {Random House New York, 1979}.

\bibitem{corcos} A. Corcos, J.-P. Eckmann, A. Malaspinas, Y. Malevergne and D. Sornette,
          Quantitative Finance {\bf 2}, 264 (2002).

\bibitem{galam} S. Galam,
        Physica A {\bf 330}, 453 (2004).

\bibitem{mobilia} M. Mobilia and S. Redner,
        Phys. Rev. E {\bf 68}, 046106 (2003).

\bibitem{challet} D. Challet and Y.C. Zhang,
        Physica A {\bf 246}, 407(1997); {\bf 256}, 514(1998).

\bibitem{arthur}W.B. Arthur,
        Amer. Econ. Assoc. Papers Proc. {\bf 84}, 406(1994).

\bibitem{web} See, for example, the website on econophysics
        at http://www.unifr.ch/econophysics/.

\bibitem{challet1}D. Challet, M. Marsili, and Y.C. Zhang,
        Physica A {\bf 294}, 514(2001); {\bf 299}, 228(2001).

\bibitem{johnson} N.F. Johnson, P. Jefferies, and P.M. Hui,
        {\it Financial Market complexity} (Oxford University Press, Oxford, 2003).

\bibitem{savit} R. Savit, R. Manuca, and R. Riolo,
        Phys. Rev. Lett. {\bf 82}, 2203(1999).

\bibitem{challet2}D. Challet, M. Marsili, and R. Zecchina,
        Phys. Rev. Lett. {\bf 84}, 1824(2000); {\bf 85}, 5008(2000).

\bibitem{zheng}D.F. Zheng and B.H. Wang,
        Physica A {\bf 301}, 560(2001).

\bibitem{challet3}D. Challet and Y.C. Zhang,
        Physica A {\bf 269}, 30(1999)

\bibitem{manuca}R. Manuca, Y. Li, R. Riolo, and R. Savit,
        Physica A {\bf 282}, 559(2000).

\bibitem{johnson1}N.F. Johnson, M. Hart, and P.M. Hui,
        Physica A {\bf 269}, 1(1999).

\bibitem{hart}M. Hart, P. Jefferies, N.F. Johnson, and P.M. Hui,
        Physica A {\bf 298}, 537(2001).

\bibitem{cavagna}A. Cavagna, J.P. Garrahan, I. Giardina, and D.
        Sherrington,
        Phys. Rev. Lett. {\bf 83}, 4429(1999).

\bibitem{johnson2}N.F. Johnson, P.M. Hui, R. Jonson, and T.S. Lo,
        Phys. Rev. Lett. {\bf 82}, 3360(1999).

\bibitem{cara}M. de Cara, O. Pla, and F. Guinea,
        Eur. Phys. J. B {\bf 13}, 413(2000).

\bibitem{li}Y. Li and R. Savit,
        Physica A {\bf 335}, 217(2004).

\bibitem{johnson3}N.F. Johnson, P.M. Hui, D.F. Zheng, and M. Hart,
        J. Phys. A: Math. Gen. {\bf 32}, L427(1999).

\bibitem{lo} T.S. Lo, H.Y. Chan, P.M. Hui, and N.F. Johnson, Phys.
Rev. E {\bf 70}, 056102 (2004).

\bibitem{sonic} H.Y. Chan, T.S. Lo, P.M. Hui, and N.F. Johnson,
e-print cond-mat/0408557.

\bibitem{larry} K.F. Yip, T.S. Lo, P.M. Hui, and N.F. Johson,
Phys. Rev. E {\bf 69}, 046120 (2004).

\end{thebibliography}
\end{document}